\documentclass[prx,floatfix,10pt,twocolumn,nofootinbib,table, dvipsnames]{revtex4-2}

\usepackage[T1]{fontenc}
\usepackage[utf8]{inputenc}
\usepackage{microtype}
\usepackage{array}
\usepackage{graphicx}
\usepackage{dcolumn}
\usepackage{bm}
\usepackage{amsmath}
\usepackage{amsfonts}
\usepackage{amssymb}
\usepackage{amstext}   
\usepackage{amsthm}
\usepackage{array}
\usepackage[english]{babel}
\usepackage{makecell}
\usepackage{dsfont}
\usepackage{nameref}
\usepackage{color}
\usepackage{float}
\usepackage[outdir=./]{epstopdf}
\usepackage{nameref}
\usepackage{subfloat}
\usepackage[colorlinks = true, citecolor = blue, urlcolor =cyan]{hyperref}
\usepackage[figure]{hypcap}
\usepackage{multirow}
\usepackage{makecell}
\usepackage{xcolor}
\usepackage{mathtools}
\usepackage{mathdots}
\usepackage[notransparent]{svg}
\usepackage{todonotes} %I like to use this one to highlight missing bits for myself. -Alicia
\newcolumntype{C}{>{$}c<{$}}
\AtBeginDocument{
\heavyrulewidth=.08em
\lightrulewidth=.05em
\cmidrulewidth=.03em
\belowrulesep=.65ex
\belowbottomsep=0pt
\aboverulesep=.4ex
\abovetopsep=0pt
\cmidrulesep=\doublerulesep
\cmidrulekern=.5em
\defaultaddspace=.5em
\tabcolsep=7pt
}
\usepackage{booktabs}
\usepackage{verbatim}

\definecolor{emerald}{rgb}{0.07, 0.53, 0.03}

\usepackage{braket}

\usepackage{nicefrac} %added for equations

\usepackage{xcolor} %%%%WARNING arxiv is very touchy about this!!!!!
\usepackage{color}
\usepackage{contour}
\newcommand{\specialstar}{{\footnotesize\contour{Red}{$\color{Goldenrod}\bigstar$}}}
\newcommand{\specialtriangle}{{\footnotesize\contour{Red}{$\color{Goldenrod}\blacktriangle$}}}

\begin{document}

\title{Realization of pure gyration in an on-chip superconducting microwave device}

\author{
\large{
{Zhiyin Tu$^{1,2}$}, {Violet Workman$^3$}, {Gaurav Bahl$^{4,5}$}, \mbox{and Alicia J. Koll\'ar$^{2,6,7,\dagger}$}} \\
\vspace{6pt}
    \footnotesize{$^1$ Department of Electrical and Computer Engineering, University of Maryland, College Park, MD 20742, USA,} \\
    \footnotesize{$^2$ Joint Quantum Institute, University of Maryland, College Park, MD 20742, USA,} \\
    \footnotesize{$^3$ Department of Physics, University of Illinois at Urbana\textendash Champaign, Urbana, IL 61801, USA,}\\
    \footnotesize{$^4$ Department of Mechanical Science and Engineering, University of Illinois at Urbana–Champaign, Urbana, IL 61801 USA} \\
    \footnotesize{$^5$ Illinois Quantum Information Science and Technology Center, University of Illinois at Urbana–Champaign, Urbana, IL 61801 USA} \\
    \footnotesize{$^6$ Department of Physics, University of Maryland, College Park, MD 20742, USA} \\
    \footnotesize{$^7$ Quantum Materials Center, University of Maryland, College Park, MD 20742, USA} \\ 
    \footnotesize{$\dagger$ Corresponding author: \href{mailto:akollar@umd.edu}{akollar@umd.edu}}
}

\maketitle

%\begin{abstract}

\bf{}
Synthetic materials that emulate tight-binding Hamiltonians have enabled a wide range of advances in topological and non-Hermitian physics. A crucial requirement in such systems is the engineering of non-reciprocal couplings and synthetic magnetic fields.
More broadly, the development of these capabilities in a manner compatible with quantum-coherent degrees of freedom remains an outstanding challenge, particularly for superconducting circuits, which are highly sensitive to magnetic fields. 
Here we demonstrate that \emph{pure gyration} -- a non-reciprocal coupling with exactly matched magnitude but non-reciprocal $\pi$ phase contrast -- can be realized between degenerate states using only spatio-temporal modulation. 
Our experiments are performed using microwave superconducting resonators that are modulated using dc-SQUID arrays.
We first show the existence of continuous exceptional surfaces in modulation parameter space where coupling with arbitrarily-large magnitude contrast can be achieved, with robust volumes of $\pi$ phase contrast contained within.
We then demonstrate that intersection of these volumes necessarily gives rise to new continuous surfaces in parameter space where pure gyration is achieved.
With this we experimentally demonstrate $>58$ dB isolation and the first on-chip gyrator with only superconducting circuit elements.
Our method is fully agnostic to physical implementation (classical or quantum) or frequency range and paves the way to large-scale non-reciprocal metamaterials.
\rm{}

%\end{abstract}

Synthetic materials that emulate tight-binding Hamiltonians have proven instrumental in major advances in condensed matter and topological physics \cite{carusotto2020}.
The generation of asymmetric couplings and effective magnetic fields in these synthetic materials has, in particular, enabled explorations of quantum Hall physics \cite{guinea2009, Cho2008, Wang2008, Koch2010, wang2009, rechtsman2013, Aidelsburger2011, Aidelsburger2013, Ozawa2016, Lohse2018}, led to the discovery and first demonstrations of higher-order topological insulators \cite{Benalcazar2017, Peterson2018, serra2018, yamada2022}, 
and the demonstrations of non-Hermitian skin effects \cite{HN1996, HN1998, brandenbourger2019, topolectricalcircuits2020, NHSE, zhang2021, liu2022, Gao2023}
and topological funneling \cite{Weidemann2020}.
A specific requirement for non-reciprocal synthetic \emph{quantum} materials -- in which non-reciprocal couplings are combined with quantum coherent degrees of freedom -- is the ability to generate the asymmetry at low energy densities in an environment compatible with quantum control and coherent atoms or two-level systems.
Superconducting circuits \cite{blais2021circuit} are a prime candidate in this context, having realized lattice metamaterials \cite{carusotto2020, houck2012, ma2019} with reciprocal couplings.

While non-reciprocal couplers can be readily implemented in room-temperature electronics, mechanics, and microwave cavities \cite{brandenbourger2019,topolectricalcircuits2020,zhang2021,salcedo2025,Clai2018}, these approaches are incompatible for on-chip integration with superconducting quantum microwave circuits.
To date, non-reciprocal coupling in superconducting circuits has only been possible between \emph{non-degenerate} states by means of time modulation \cite{roushan2017, rosen2024, Lecocq2017, rosen2025}, such that the couplings are complex and each direction imparts equal and opposite tunable phase with no magnitude asymmetry. Couplings of this kind implement synthetic magnetic fields.
In related work in photonics, spatio-temporal modulation has been shown to generate non-reciprocal coupling between \emph{degenerate} optical states \cite{Orsel2025}. This technique allows couplings with arbitrarily large magnitude contrast, and additionally enables gyration, i.e., coupling with $\pi$ phase contrast. 
It is therefore particularly appealing for superconducting circuits as similar modulations could be implemented with relative ease.
\emph{Pure gyration}, in which a non-reciprocal $\pi$ phase contrast is accompanied by perfect magnitude symmetry, is the fundamental building block of non-reciprocal signal processing devices \cite{pozar2011}.  

%%%%normal figure
\begin{figure*}[t]
\centering
    \includegraphics[width=1\textwidth]{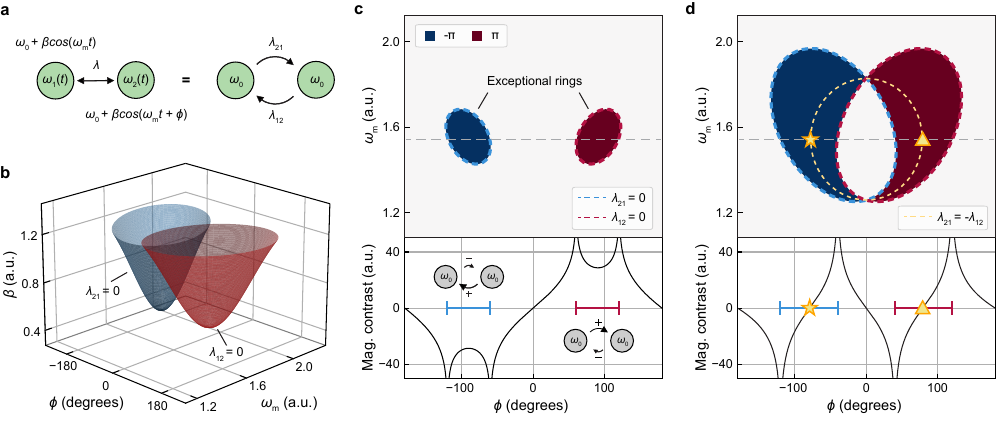}
    \vspace{-0.2cm}
	\caption{\label{fig1}
            \textbf{Origin of pure gyration coupling through spatio-temporal modulation.}
            \textbf{a}, A pair of degenerate states with time modulated on-site energy can be shown to map to an equivalent static Hatano-Nelson (HN) model \cite{Orsel2025} that has $\lambda_{21} \neq \lambda_{12}$ when $\phi \neq 0^\circ, \pm180^\circ$. An exceptional point (EP) is obtained when either $\lambda_{21}$ or $\lambda_{12}$ goes to zero.
            \textbf{b}, The 3D parameter space defined by $\beta$, $\omega_\text{m}$, and $\phi$ reveals two exceptional surfaces corresponding to zeros of either coupling.
            \textbf{c}, A plane cut of \textbf{b} taken at moderate $\beta$ shows independent exceptional rings, inside which exactly one of the couplings becomes negative, indicating a relative $\pi$ phase contrast. 
            Log-scale plot of the coupling magnitude contrast along the gray dashed line shows zeros only at $\phi=0^\circ,\pm180^\circ$ and singularities at the EPs.
            \textbf{d}, Above a critical modulation strength the exceptional rings intersect, resulting in an alternation of the EPs along the cut indicated by the gray dashed line. 
            This forces additional zeros in the magnitude contrast at non-trivial $\phi$ indicated by \protect\specialstar\ and \protect\specialtriangle, where $\lambda_{21} = -\lambda_{12}$ (pure gyration) must occur. 
            The gold dashed trajectory indicates the guaranteed existence of this pure gyration behavior over the entire $\omega_\text{m}$ range where the intersection occurs.}  
\end{figure*}

In this work, we demonstrate that pure gyration coupling between a pair of degenerate states can be produced through only spatio-temporal modulation. 
Our experimental system is comprised of a pair of degenerate superconducting microwave resonators, each of which contains a dedicated frequency modulator based on dc superconducting quantum interference devices (dc-SQUIDs) \cite{tinkham2004introduction}.
First, we go far beyond the exploration performed in photonics \cite{Orsel2025} and reveal a larger parameter space that supports exceptional point (EP) surfaces where giant coupling contrast can be achieved.
Using this we demonstrate $>58$ dB microwave isolation, which is among the highest values reported in a chip-scale superconducting microwave device \cite{IBM2023wideband,Lehnert2017widely,Murch2024superconducting,cao2024parametrically,bernier2017nonreciprocal,Lecocq2017}.
We then show that beyond these EPs lie islands of robustness where the phase non-reciprocity of the coupling holds perfectly at $\pi$ and is extremely insensitive to parameter choice.
And most excitingly, we find that once a critical modulation threshold is surpassed, these islands can intersect, and as a result, pure gyration coupling, with $\pi$ phase non-reciprocity and exactly matched magnitude, can be found. The pure gyration effect is not an isolated singularity point and is also shown to be robust to parameter choice.
When we use this coupling in a one-dimensional chain with an even number of sites, e.g., with 2 superconducting resonators in this study, the transmission response is that of a gyrator.         
This work reveals new regimes of operation for spatio-temporally modulated states that are fully agnostic to implementation and should be accessible in any classical or quantum system. 
Our experimental demonstration is uniquely aided by the high dynamic range of the SQUID modulators, and uses \emph{only} superconducting circuit components compatible with superconducting-qubit quantum computing and no off-chip elements. As a result, these results constitute the first ever fully on-chip demonstration of pure gyration for quantum-microwave signals \cite{IBM2017gyrator,RTgyrator,STMgyrator}.

\vspace{10pt}

\begin{figure*}[t]
\centering
    \includegraphics[width=1\textwidth]{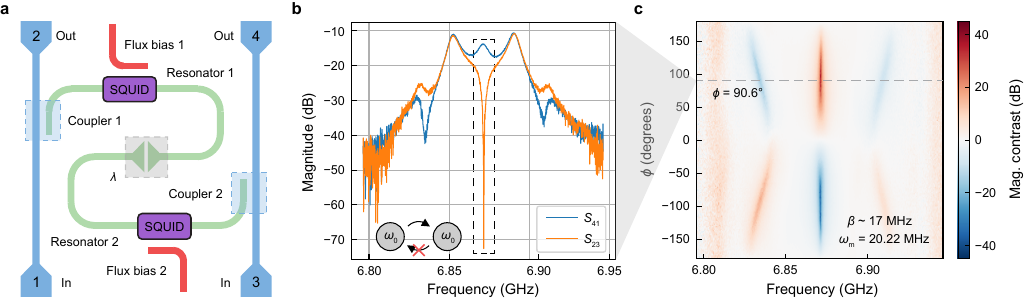}
    \vspace{-0.2cm}
	\caption{\label{fig2}
    \textbf{Device schematic and measurement of giant magnitude contrast near an exceptional point.} 
    \textbf{a}, Schematic of our experimental chip, composed of two degenerate superconducting resonators (nominal $\omega_0 = 6.8705$~GHz) with dc-SQUID frequency modulators that are controlled by independent flux bias lines. The capacitively coupled feedlines provide four input/output ports. 
    Forward through-chain transmission is measured via $S_{41}$ (backward transmission via $S_{23}$).
    Single-sided transmission measurements ($S_{21}$ and $S_{43}$) are used for individual resonator characterization and device calibration (see {\S}IV in Supplementary Information for details). 
    \textbf{b}, Experimental measurements of forward and backward transmission demonstrating the system response near an EP, confirming $\lambda_\text{12} \rightarrow 0$ through observation of $\approx58.6$~dB giant magnitude contrast at $\omega_0$.
    Here we set $\beta\sim17$~MHz, $\omega_\text{m} = 20.22$~MHz, and $\phi=90.6^{\circ}$, which is close to the bottom tip of the surfaces in Fig~\ref{fig1}b.  
    \textbf{c}, 
    A map of the magnitude contrast spectrum (defined as $|S_{41}|-|S_{23}|$) vs differential modulation phase $\phi$, with color indicating the directionality.  
    At $\omega_0$, magnitude contrast exceeding $20$ dB can be found over an $85^\circ$ range of $\phi$ (similarly for negative $\phi$ with negative contrast).
            }  
\end{figure*}

We consider a simple case of two degenerate states with reciprocal mutual coupling $\lambda$ (Fig.~\ref{fig1}a). The on-site energies are temporally modulated such that $\omega_1 = \omega_0 + \beta\cos(\omega_{\text{m}} t)$, and $\omega_2 = \omega_0 + \beta\cos(\omega_{\text{m}} t + \phi)$, where $\beta$ is the modulation strength, $\omega_{\text{m}}$ is the modulation frequency, and $\phi$ is the differential modulation phase. 
In the analytically tractable case with only one upper and one lower sideband, it can be readily shown that, for fixed $\beta$ and $\omega_\text{m}$, $\phi = \pm90^\circ$ maximizes the non-reciprocal transmission through the short chain \cite{peterson2019strong}. More recently, it was shown that this system directly maps to an equivalent \emph{static} case of two states with directional asymmetry in the linear response coupling coefficient at $\omega_0$ producing $\lambda_{21} \neq \lambda_{12}$, even with an arbitrary number of sidebands, and thus implements the generalized Hatano-Nelson (HN) model \cite{Orsel2025}. The couplings can be described in the form $\lambda_{21, 12} = \lambda(1 \pm \alpha)$ where real valued $\alpha (\beta, \omega_\text{m})$ is the degree of asymmetry that increases with larger $\beta$.

Qualitatively, the temporal modulation of the on-site energies generates a synthetic dimension in the frequency domain, which together with the physical axis of the chain, forms a two-dimensional synthetic lattice~\cite{peterson2019strong, Orsel2025} (see {\S}I and {\S}II in Supplementary Information for details).
Due to the phase-staggering nature of the modulation, a synthetic magnetic field is generated within each plaquette of the synthetic lattice. It is interference between coupling paths involving many higher-order sidebands that gives rise to the equivalent non-Hermitian HN couplings \cite{Orsel2025}. 
For fixed $\phi = \pm90^\circ$ there exist arcs of EPs in the space mapped by $\beta$ and $\omega_\text{m}$ where $|\alpha| \rightarrow 1$ and one of the coupling coefficients $\lambda_{21, 12}$ becomes exactly zero \cite{Orsel2025}. 
Beyond these arcs, we find $|\alpha|>1$, which causes a sign inversion in the corresponding $\lambda_{21, 12}$, producing in a system with gyration (non-reciprocity in phase) accompanied by finite magnitude contrast.
From these previous studies, there seemed to be no possibility of the pure gyration case as no value of $\alpha$ can produce $\lambda_\text{21}=-\lambda_\text{12}$.

In this study we permit $\phi$ to take any value and therefore generalize $\lambda_{21, 12} = \lambda (1 - \alpha_{21,12})$ where $\alpha_{21,12} (\beta, \omega_\text{m}, \phi)$ are real-valued (see Supplementary Information {\S}II for details).
This reveals continuous cone-like exceptional surfaces within the three-dimensional space mapped by $\beta$, $\omega_\text{m}$, and $\phi$ (Fig.~\ref{fig1}b) that have not previously been considered.
For moderate $\beta$, the EP first appears at $\phi=\pm90^\circ$, which corresponds to the tips of the two cone-like disjoint exceptional surfaces and the cases discussed in \cite{peterson2019strong, Orsel2025}.
As $\beta$ increases, rings of EPs emerge in the $\omega_\text{m}$-$\phi$ planes (Fig.~\ref{fig1}c), corresponding to points where either of $\lambda_{21,12} \rightarrow 0$. 
Within these rings the corresponding $\lambda_\text{21,12}$ becomes negative, implying a non-reciprocal coupling with exactly $\pi$ phase contrast.
Under these circumstances, the magnitude contrast (defined as $20\log_{10}\left| \lambda_{21}/\lambda_{12}\right|$) \emph{only} vanishes at trivial phases $\phi=0^\circ,\pm180^\circ$; whereas the $\pi$ phase contrast only occurs between the EP singularities in other ranges of $\phi$. As a result, $\pi$ phase contrast is \emph{always} accompanied by magnitude contrast.

Above a critical modulation strength, however, the exceptional surfaces enlarge and intersect (Fig~\ref{fig1}b,d), which is the key to achieving pure gyration coupling.
Over the entire range of $\omega_\text{m}$ where the surfaces intersect, we now find an alternation of the EPs contributed by either ring (Fig~\ref{fig1}d, bottom), which, due to the continuity of the magnitude contrast function, must force zeros in the contrast at non-trivial $\phi$ within the volume of influence of exactly one of the exceptional surfaces.
Thus, \emph{pure gyration} is guaranteed to appear over a continuous surface outside the intersecting volume.
Since the pure gyration effect is required by continuity, it is not an isolated fine-tuned point, and the existence of pure gyration is robust to parameter choice.
In the experiments that follow, we work with a two-site chain, and the transmission properties of the system at $\omega_0$ are directly inherited from the effective couplings. We therefore refer to the couplings and the transmission characteristics interchangeably.

\vspace{12pt}

%%%%normal figure
\begin{figure*}[t]
\centering
    \includegraphics[width=1\textwidth]{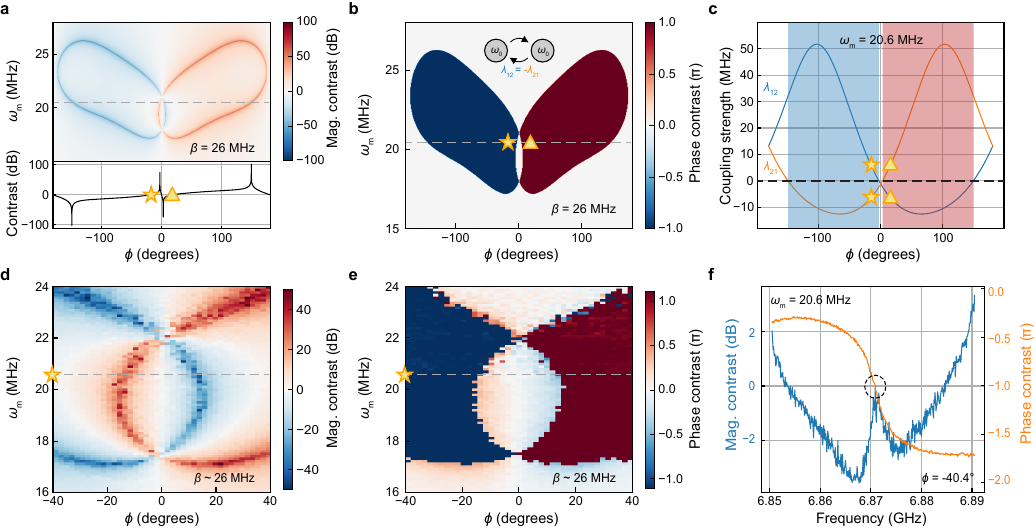}
    \vspace{-0.2cm}
	\caption{\label{fig3}
    \textbf{Experimental realization of pure gyration coupling.}
        \textbf{a}, This numerical simulation of the magnitude contrast incorporates the experimentally measured properties of our test device to confirm that $\beta = 26$ MHz is above the critical modulation strength at which the exceptional rings (corresponding to either $\lambda_{21} = 0$ or $\lambda_{12} = 0$) intersect.
        \protect\specialstar\ and \protect\specialtriangle\ point out the new zeros in the magnitude contrast function for $\omega_\text{m} = 20.6$ MHz.
        \textbf{b}, The simulation predicts that these magnitude contrast zeros sit firmly within regions of $\pi$ phase contrast.
        \textbf{c}, We plot the calculated HN coupling coefficients $\lambda_{21, 12}$ for the line cut at $\omega_\text{m} = 20.6$~MHz, confirming that at either \protect\specialstar\ or \protect\specialtriangle\ we have $\lambda_{21} = -\lambda_{12}$. Blue and red shading indicates regions where $\lambda_{21}$ and  $\lambda_{12}$ have opposite sign.
        \textbf{d,e}, Experimental measurements of magnitude contrast confirm the crossing of the exceptional rings, and the robust regions of $\pi$ phase contrast within the rings. In \textbf{d} the regions in white indicate continuous lines on which there is no magnitude contrast. The white arcs that lie within the $\pi$ phase contrast regions exhibit pure gyration.
        \textbf{f}, A specific experimental measurement corresponding to \protect\specialstar\ ($\omega_\text{m}=20.6$ MHz and $\phi=-40.4^\circ$) confirms the pure gyration effect where the magnitude contrast goes to zero while the phase contrast goes to $\pi$ at $\omega_0$ (highlighted in the black dashed circle).
            }  
\end{figure*}

For experimental implementation we developed a superconducting microwave device consisting of two frequency-tunable resonators, as shown schematically in Fig.~\ref{fig2}a. 
The nominally identical coplanar waveguide (CPW) transmission line resonators are capacitively coupled with static $\lambda\approx16.4$~MHz. The tunability of the resonance frequency comes from a dc-SQUID array embedded in the middle of each resonator. Each SQUID serves as a flux-tunable inductance with 
    \begin{equation}
        L_{\text{SQUID}}(\Phi_e) = \frac{\hbar}{2e I_c \sqrt{1 + \gamma^2 + 2\gamma \cos(2\pi\frac{\Phi_e}{\Phi_0})}},
    \label{eq2}
    \end{equation}
where $\gamma$ is the asymmetry factor between the two Josephson junctions that form the SQUID, $I_c$ is the critical current for the smaller junction, $\Phi_e$ is the external magnetic flux threading the loop, and $\Phi_0=h/2e$ is the superconducting magnetic flux quantum \cite{switch2016, quantumengineering}. By putting $N$ identical SQUIDs in series and forming an array to achieve a certain total inductance $L$, the required inductance of each single SQUID is reduced, improving the linearity of the device (1-dB compression point) \cite{switch2016}. Here we choose $N=2$ for better flux homogeneity. We supply both ac and dc 
bias voltages through the individual flux bias line next to the SQUID arrays such that the frequency of the resonators has a nominal value of $\omega_0=6.8705$~GHz with a superimposed sinusoidal modulation at $\omega_\text{m}$.
The device fabrication, measurement setup, and in-situ calibrations are discussed in Supplementary Information {\S}III and {\S}IV.

The device features two capacitively-coupled feedlines, one coupled to each resonator, giving rise to four input and output ports, as shown in Fig.~\ref{fig2}a. Transmission from port 1 to port 4 ($S_{41}$) and from port 3 to port 2 ($S_{23}$) pass through the resonator chain in the forward and backward directions, respectively. Differences between $S_{41}$ and $S_{23}$ (after in-situ calibration, see Fig. S7 in Supplementary Information for details) provide evidence for the non-reciprocity in the effective HN coupling coefficients. Single-sided transmission measurements ($S_{21}$ and $S_{43}$) are used for individual resonator characterization and modulation calibration (see {\S}IV in Supplementary Information for details). 

\vspace{12pt} 

We first confirm the existence of an EP near $\phi = \pm 90^\circ$ for moderate modulation strength, where strong non-reciprocal behavior is expected \cite{peterson2019strong}. The measurements in Fig.~\ref{fig2}b, with modulation frequency $\omega_{\text{m}}=20.22$~MHz and modulation strength $\beta \sim 17$ MHz, confirm that an extremely large transmission magnitude contrast ($>58$ dB) can be obtained at $\omega_0$ when one of the HN couplings approaches zero. 
In Fig.~\ref{fig2}c we present additional measurements of the magnitude contrast over a large sweep of $\phi$. These measurements confirm the robustness of the non-reciprocal behavior to the choice of modulation parameters. This data, taken along a line of fixed $\omega_\text{m}$ and $\beta$ (with variable $\phi$) which is tangent to the exceptional surfaces, shows magnitude contrast at $\omega_0$ in excess of $20$ dB over an $85^\circ$ range of $\phi$ (similarly for negative $\phi$ with negative contrast).

In Fig.~\ref{fig3}a-c we present numerical simulations of the device response at $\omega_0$ for stronger $\beta$. Here we find that $\beta \sim 26$ MHz is above the threshold modulation strength at which intersecting rings of EPs (corresponding to either $\lambda_{21} = 0$ or $\lambda_{12} = 0$) appear in the $\omega_\text{m}$-$\phi$ plane.
As predicted from the theoretical discussion in Fig.~\ref{fig1}, we can identify points of interest for any choice of $\omega_\text{m}$ within the intersection range, $17.5$\textendash$22$ MHz, where magnitude contrast must go to zero while the phase contrast stays robustly at $\pi$. In Fig.~\ref{fig3}c we select a slice at $\omega_\text{m} = 20.6$ MHz to identify these points of pure gyration HN coupling ($\lambda_{21} = -\lambda_{12}$) at non-trivial values of the differential modulation phase $\phi$.

In Fig.~\ref{fig3}d,e we present experimental measurements of the magnitude contrast and phase contrast at the intersection region. Continuity of the HN couplings and the magnitude contrast enforce the existence of a pure gyration point for \emph{every} modulation frequency over the intersection since zeros of the magnitude contrast (white arcs in Fig.~\ref{fig3}d) lie fully inside the $\pi$ phase contrast regions (Fig.~\ref{fig3}e) resulting from the rings of EPs. In Fig~\ref{fig3}f we present transmission measurements at a specific operating point indicated by \specialstar, where $\omega_{\text{m}}=20.6$~MHz and $\phi=-40.4^{\circ}$. As expected, at $\omega_0$ the measured phase contrast of $\pi$ coincides with a near-zero magnitude contrast.

\vspace{12pt}

The method presented here of achieving pure gyration and non-reciprocal couplings between degenerate degrees of freedom via spatio-temporal modulation is both simple and versatile, and can be extended to signal and information processing applications in \emph{both} the classical and quantum regimes.
The 
sign change of the coupling coefficients $\lambda_{21,12}$ when crossing the exceptional surfaces leads to a switchable $\pi$ phase shift in transmission, which could be used to implement an extremely robust phase-shift keying protocol for classical telecommunications in which the phase shift is intrinsically protected, and control errors affect only the signal magnitude.

In the quantum regime, quantum computers based on superconducting qubits \cite{blais2021circuit,aumentado2020parametric,krinner2019} rely on circulators and isolators to protect qubits from thermal radiation and amplifier backaction \cite{aumentado2020parametric}. 
However, the current ferrite-based solutions are incompatible with on-chip integration due to their bulky form factor, which is orders of magnitude larger than the quantum processors they facilitate \cite{krinner2019}.
The device presented here, implemented using superconducting resonators in a dilution refrigerator, is compatible with superconducting qubits and
adds entirely new capabilities to the toolbox of on-chip non-reciprocal devices for superconducting circuits \cite{Murch2024superconducting, Lecocq2017, IBM2023wideband,Lehnert2017widely,cao2024parametrically,bernier2017nonreciprocal,IBM2017gyrator}. In addition to being the first ever fully on-chip realization of a gyrator for superconducting microwave devices, our device also realizes extremely high isolation ($>58$ dB), as well as the ability to dynamically reconfigure the direction of isolation. 
The combination of strong non-reciprocity and in-situ reconfigurability makes this method ideal for narrower-band tasks such as shielding qubits from byproducts of parametric amplifiers \cite{aumentado2020parametric} or directional routing of quantum signals \cite{cao2024parametrically,directionalrouting}.
Furthermore, since superconducting transmon qubits \cite{blais2021circuit} are formed using anharmonic oscillators, the simple spatio-temporal modulation method described here can also be employed between two degenerate qubits, rather than two resonators, where it should lead to non-reciprocal parametric two-qubit interactions.

%%%%%%%%%%%%%%%%%%%%%%%%%%%%%%%%%%%
\vspace{2cm} 

\bibliographystyle{apsrev4-2}
% {\footnotesize \bibliography{main_refs.bib}}
\bibliography{main_refs.bib}

\vskip 0.2in\vskip 0.2in
\section*{Acknowledgements}
We thank Ashish Clerk, Oded Zilberberg, Andrew Daley, Penghao Zhu, Taylor Hughes, Jacopo Gliozzi, and Thomas Antonsen for helpful conversations and comments.

This work was supported by grants No. W911NF-23-1-0219 (ARO), No. W911NF-17-S-0003 (ARL), OMA-2120757 (NSF QLCI), and N00014-20-1-2325 (ONR MURI).

\section*{Author Contributions}
A.J.K. and G.B. conceived of the work and oversaw the project. V.W. assisted with numerical simulations and calculation of effective HN parameters. Z.T. designed and fabricated the device and carried out all measurements and led the analysis with contributions from all authors. All authors jointly wrote the manuscript.

\section*{Data Availability}
Correspondence and requests for materials should be addressed to A.J.K (email: akollar@umd.edu). 
The data are available from the corresponding author upon reasonable request.

\end{document}

\title{Supplementary Information for:\\Realization of pure gyration in an on-chip superconducting microwave device}

\author{
\large{
{Zhiyin Tu$^{1,2}$}, {Violet Workman$^3$}, {Gaurav Bahl$^{4,5}$}, \mbox{and Alicia J. Koll\'ar$^{2,6,7,\dagger}$}} \\
\vspace{6pt}
    \footnotesize{$^1$ Department of Electrical and Computer Engineering, University of Maryland, College Park, MD 20742, USA,} \\
    \footnotesize{$^2$ Joint Quantum Institute, University of Maryland, College Park, MD 20742, USA,} \\
    \footnotesize{$^3$ Department of Physics, University of Illinois at Urbana\textendash Champaign, Urbana, IL 61801, USA,}\\
    \footnotesize{$^4$ Department of Mechanical Science and Engineering, University of Illinois at Urbana–Champaign, Urbana, IL 61801 USA} \\
    \footnotesize{$^5$ Illinois Quantum Information Science and Technology Center, University of Illinois at Urbana–Champaign, Urbana, IL 61801 USA} \\
    \footnotesize{$^6$ Department of Physics, University of Maryland, College Park, MD 20742, USA} \\
    \footnotesize{$^7$ Quantum Materials Center, University of Maryland, College Park, MD 20742, USA} \\ 
    \footnotesize{$\dagger$ Corresponding author: \href{mailto:akollar@umd.edu}{akollar@umd.edu}}
}
% \author{Zhiyin Tu}
% \affiliation{Department of Electrical and Computer Engineering and JQI, University of Maryland, College Park, MD 20742, USA}

% \author{Violet Workman}
% \affiliation{Department of Physics, University of Illinois at Urbana–Champaign, Urbana, IL 61801, USA}

% \author{Gaurav Bahl}
% \affiliation{Department of Mechanical Science and Engineering, University of Illinois at Urbana–Champaign, Urbana, IL 61801, USA}

% \author{Alicia J.  Koll\'ar}
% \email{akollar@umd.edu}
% \affiliation{Department of Physics and JQI, University of Maryland, College Park, MD 20742, USA}

% \preprint{APS/123-QED}

% \date{\today}

% \begin{abstract}
% Blah blah blah asbtract. lakjsdhflkajshdlkfjahsdlkjfhalskjdhflaksjdh
% \end{abstract}

%\keywords{Suggested keywords}%Use showkeys class option if keyword
                              %display desired
\maketitle

%\tableofcontents

\raggedbottom

\section{\label{sup:input-output}Input-Output Relations of a Spatio-temporally Modulated Resonator Chain}

\begin{figure}[!htbp]
\centering
        \includegraphics{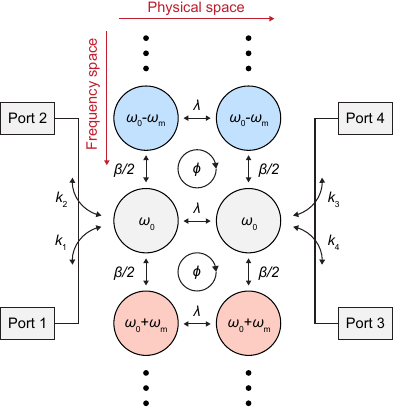}
	\caption{\textbf{Schematic of a time-invariant 2D lattice with a synthetic dimension in frequency.} Two feedlines coupled to the ends of the resonator chain provide 4 ports for inputs and outputs. $k_1$ to $k_4$ represent the coupling coefficients to each port, respectively. Modulation strength $\beta$ sets the couplings between sites in the synthetic dimension, and the differential modulation phase $\phi$ generates non-reciprocity by imprinting direction-dependent phase shifts onto the couplings. All the contributions to the transmission at $\omega_0$ through sidebands can be effectively collapsed into the asymmetric Hatano-Nelson couplings $\lambda_{21}$ and $\lambda_{12}$ as shown in Fig. 1a in the main manuscript.
}
	\label{Fig. S1}
\end{figure}

We use temporal coupled mode theory to derive the input-output relations of the modulated coupled resonator chain studied in this work. This type of treatment for modulated resonator networks was first introduced in Refs \cite{suh2004temporal, peterson2019strong, Orsel2025}. Here we provide a complete derivation of the input-output relations of our system with modification to adapt to our specific case of hanger-style resonator-feedline coupling. The overall structure of the calculation is identical to that shown in previous works, but accurate simulation of the scattering matrix requires incorporating the non-chiral nature of the hanger-style coupling in superconducting microwave circuits, which differs from the typical chiral couplers in photonics, and necessitates a full 4-port description. 

The resonance frequencies of the two resonators are modulated sinusoidally such that
\begin{equation}
\begin{aligned}
\omega_1(t) &= \omega_0 + \beta \cos(\omega_{\text{m}}t), \\
\omega_2(t) &= \omega_0 + \beta \cos(\omega_{\text{m}}t + \phi).
\end{aligned}
\label{eq_S1}
\end{equation}
The excitation amplitudes of resonator 1 and 2 can be written collectively as $\ket{a(t)}=[a_1(t), a_2(t)]^T$. 
% \ak{An input-output theory treatment for the transmission properties and effective HN coefficients of such modulated resonator chains was worked out in detail in Refs. XYZ. Here we provide a summary of their treatment modified to the specific case of hanger-style resonator-feedline coupling used here. The structure of the calculation is identical to that shown in these previous works, but accurate simulation of S parameters requires incorporating the non-chiral nature of hanger-style coupling.}
Following Refs. \cite{suh2004temporal, peterson2019strong, Orsel2025}, the dynamic equation for the time evolution of $\ket{a(t)}$ is
\begin{equation}
\frac{\partial}{\partial t}\ket{a(t)}=[i\Omega_0+i\Omega_1(t)-\Gamma]\ket{a(t)} + K^T\ket{s_+(t)},
\label{eq_S2}
\end{equation}
where
\begin{equation}
\begin{aligned}
\Omega_0 &= \begin{pmatrix}
\omega_0 & \lambda \\
\lambda & \omega_0
\end{pmatrix}, \quad 
\Gamma = \begin{pmatrix}
\gamma & 0 \\
0 & \gamma
\end{pmatrix}, \quad 
K = \begin{pmatrix}
    k_1 & 0 \\
    k_2 & 0 \\
    0 & k_3 \\
    0 & k_4
\end{pmatrix}, \\
\Omega_1(t) &= \begin{pmatrix}
\beta \cos(\omega_{\text{m}}t) & 0 \\
0 & \beta \cos(\omega_{\text{m}}t + \phi)
\end{pmatrix}.
\end{aligned}
\label{eq_S3}
\end{equation}
The total decay rates are described by $\Gamma$, the couplings between the ports and resonators (shown in Fig. \ref{Fig. S1}) are described by $K$, and the incoming waves from all four ports are described by $\ket{s_+(t)}=[s_{+,1}(t), s_{+,2}(t), s_{+,3}(t), s_{+,4}(t)]^T$. As a direct consequence of energy conservation, $2\Gamma=K^\dagger K+\kappa$ holds true \cite{suh2004temporal}, where $\kappa=\text{diag}(\kappa_1, \kappa_2)$ is the internal loss matrix (note that despite conventionally being referred to as the total decay rate, $\gamma$ actually represents half the sum of the internal and external loss rates). Following Ref. \cite{manolatou1999coupling}, the outgoing waves of the system are
\begin{equation}
\begin{aligned}
s_{-,1}(t) &= s_{+,2}(t)-k^*_1 a_1(t), \\
s_{-,2}(t) &= s_{+,1}(t)-k^*_2 a_1(t), \\
s_{-,3}(t) &= s_{+,4}(t)-k^*_3 a_2(t), \\
s_{-,4}(t) &= s_{+,3}(t)-k^*_4 a_2(t),
\end{aligned}
\label{eq_S4}
\end{equation}
which can be simplified into
\begin{equation}
\ket{s_-(t)}=C\ket{s_+(t)}+D\ket{a(t)}
\label{eq_S5},
\end{equation}
where
\begin{equation}
\begin{aligned}
C = \begin{pmatrix}
0 & 1 & 0 & 0 \\
1 & 0 & 0 & 0 \\
0 & 0 & 0 & 1 \\
0 & 0 & 1 & 0
\end{pmatrix}, \quad 
D = \begin{pmatrix}
    -k^*_1 & 0 \\
    -k^*_2 & 0 \\
    0 & -k^*_3 \\
    0 & -k^*_4 
\end{pmatrix},
\end{aligned}
\label{eq_S6}
\end{equation}
and $\ket{s_-(t)}=[s_{-,1}(t), s_{-,2}(t), s_{-,3}(t), s_{-,4}(t)]^T$.

Given the periodicity of the system in time and our interest in its spectral response, it is more convenient to work in the frequency domain. After performing a Fourier transform on Eq. \ref{eq_S2} ($\ket{a(\omega)}=\int \ket{a(t)}e^{-i\omega t} \, dt$), we have
\begin{multline}
\omega\ket{a(\omega)}=H_0\ket{a(\omega)}-i K^T\ket{s_+(\omega)}\\
+B\ket{a(\omega-\omega_\text{m})}+B^\dag\ket{a(\omega+\omega_\text{m})},
\label{eq_S7}
\end{multline}
where $H_0=\Omega_0+i\Gamma$ is the Hamiltonian of the system without any modulation, and
\begin{equation}
B = \frac{\beta}{2}\begin{pmatrix}
    1 & 0 \\
    0 & e^{i\phi}
\end{pmatrix}
\label{eq_S8}
\end{equation}
is the coupling matrix between adjacent sidebands (separated by $\omega_\text{m}$) in the frequency space generated by the temporal modulation (see Fig. \ref{Fig. S1}). Since this recursive relation in Eq. \ref{eq_S7} applies to the infinitely many sidebands generated, we can extend it to the form
\begin{equation}
\omega\ket{\alpha(\omega)}=\mathcal{H}\ket{\alpha(\omega)}-i\mathcal{K}^T\ket{\sigma_+(\omega)},
\label{eq_S9}
\end{equation}
where
\begin{equation}
\mathcal{K}=\begin{pmatrix}
    \ddots & 0 & 0 & 0 & 0 \\
    0 & K & 0 & 0 & 0 \\
    0 & 0 & K & 0 & 0 \\
    0 & 0 & 0 & K & 0 \\
    0 & 0 & 0 & 0 & \ddots
\end{pmatrix},
\label{eq_S10}
\end{equation}
and
\begin{equation}
\mathcal{H}=\begin{pmatrix}
    \ddots & \ddots & 0 & 0 & 0 \\
    \ddots & H_0-\omega_\text{m}I & B & 0 & 0 \\
    0 & B^\dag & H_0 & B & 0 \\
    0 & 0 & B^\dag & H_0+\omega_\text{m}I & \ddots \\
    0 & 0 & 0 & \ddots & \ddots
\end{pmatrix}
\label{eq_S11}
\end{equation}
are both block matrices. The collective excitation amplitudes $\ket{\alpha(\omega)}$ and system input/output $\ket{\sigma_{\pm}(\omega)}$ over all sideband frequencies are
\begin{equation}
\begin{aligned}
\ket{\alpha(\omega)} = \begin{pmatrix}
\vdots \\
\ket{a(\omega+\omega_\text{m})} \\
\ket{a(\omega)} \\
\ket{a(\omega-\omega_\text{m})} \\
\vdots
\end{pmatrix}, \ket{\sigma_{\pm}(\omega)} = \begin{pmatrix}
    \vdots \\
\ket{s_{\pm}(\omega+\omega_\text{m})} \\
\ket{s_{\pm}(\omega)} \\
\ket{s_{\pm}(\omega-\omega_\text{m})} \\
\vdots 
\end{pmatrix}.
\end{aligned}
\label{eq_S12}
\end{equation}
Similarly, the output of the system at frequency $\omega$ can be derived by performing Fourier transform of Eq. \ref{eq_S5}
\begin{equation}
\ket{s_-(\omega)}=C\ket{s_+(\omega)}+D\ket{a(\omega)},
\label{eq_S13}
\end{equation}
which can again be extended to all sideband frequencies as
\begin{equation}
\ket{\sigma_-(\omega)}=\mathcal{C}\ket{\sigma_+(\omega)}+\mathcal{D}\ket{\alpha(\omega)}
\label{eq_S14}
\end{equation}
where
\begin{equation}
\begin{aligned}
\mathcal{C} = \begin{pmatrix}
    \ddots & 0 & 0 & 0 & 0 \\
    0 & C & 0 & 0 & 0 \\
    0 & 0 & C & 0 & 0 \\
    0 & 0 & 0 & C & 0 \\
    0 & 0 & 0 & 0 & \ddots
\end{pmatrix}, \quad 
\mathcal{D} = \begin{pmatrix}
    \ddots & 0 & 0 & 0 & 0 \\
    0 & D & 0 & 0 & 0 \\
    0 & 0 & D & 0 & 0 \\
    0 & 0 & 0 & D & 0 \\
    0 & 0 & 0 & 0 & \ddots
\end{pmatrix}.
\end{aligned}
\label{eq_S15}
\end{equation}
are both block matrices.

In principle, the input-output relation of the system can be found by combining Eqs. \ref{eq_S9} and \ref{eq_S14} to eliminate $\ket{\alpha(\omega)}$ and solve for the scattering matrix $\mathcal{S}(w)$ as $\ket{\sigma_-(\omega)}=\mathcal{S}(\omega)\ket{\sigma_+(\omega)}$. However, it is often more practical and useful to focus on the linear scattering matrix $S(\omega)$ that connects the input and output at the same frequency, i.e., $\ket{s_-(\omega)}=S(\omega)\ket{s_+(\omega)}$. This can be achieved by first applying the recursive relation from Eq. \ref{eq_S7} and selecting an appropriate cut-off order, $P$, for the sidebands. This allows $\ket{a(\omega-\omega_\text{m})}$ and $\ket{a(\omega+\omega_\text{m})}$ to be expressed in terms of $\ket{a(\omega)}$. With this, the relation between $a(\omega)$ and $\ket{s_+(\omega)}$ can be derived from Eq. \ref{eq_S7} and substituted back into Eq. \ref{eq_S13} to eliminate $\ket{a(\omega)}$. By doing so, we can derive the scattering matrix $S(\omega)$ as
\begin{equation}
S(\omega)=C-iD[\omega I-H_0-O_{\text{-P}}-\Tilde{O}_\text{P}]^{-1}K^T,
\label{eq_S16}
\end{equation}
where
\begin{equation}
\begin{aligned}
    O_{\text{-P}} &= B\{(\omega I - H_{-1}) - B[(\omega I - H_{-2}) - \dots \\ 
    &\quad - B[(\omega I - H_{\text{-P+1}}) - B[\omega I - H_{\text{-P}}]^{-1} B^\dag]^{-1} \\ 
    &\quad B^\dag \dots ]^{-1} B^\dag\}^{-1} B^\dag, \\[10pt]
    \Tilde{O}_{\text{P}} &= B^\dag\{(\omega I - H_{1}) - B^\dag[(\omega I - H_{2}) - \dots \\ 
    &\quad - B^\dag[(\omega I - H_{\text{P-1}}) - B^\dag[\omega I - H_{\text{P}}]^{-1} B]^{-1} \\ 
    &\quad B \dots ]^{-1} B\}^{-1} B.
\end{aligned}
\label{eq_S17}
\end{equation}
Note here we define $H_\text{p}\equiv H_0-p\omega_\text{m}I$, as the diagonal entries in Eq. \ref{eq_S11}.

\begin{figure*}[th!]
\centering

        \includegraphics{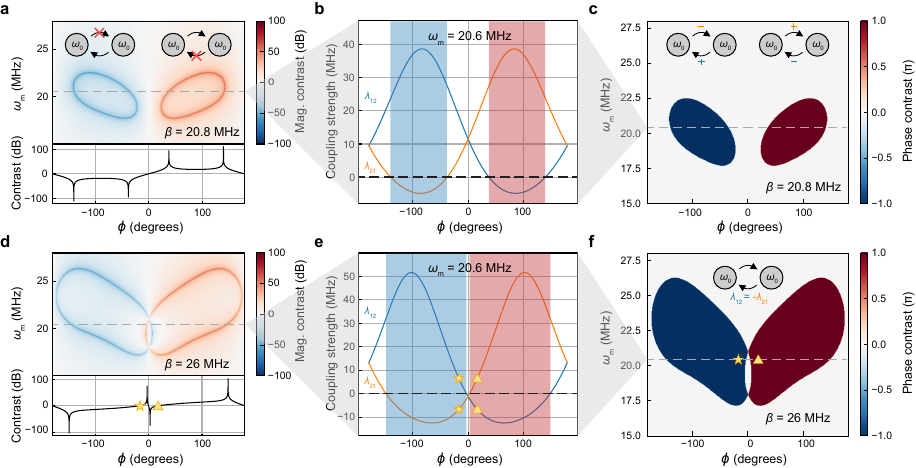}
	\caption{\textbf{Numerical simulations of two qualitatively different modulation regimes.} Here we show simulations of the phenomenology described in Fig. 1 in the main manuscript with our particular device parameters. 
  \textbf{a}-\textbf{c} show an example configuration below the critical modulation strength with $\beta=20.8$~MHz. In this regime, high transmission magnitude contrast occurs for a continuous set of modulation frequencies and phases, forming two disjoint rings in the $\omega_{\text{m}}$-$\phi$ plane (\textbf{a}), where either $\lambda_{21} = 0$ or $\lambda_{12} = 0$ (EPs). Insets indicate which coupling vanishes for positive or negative magnitude contrast, respectively. A line-cut of magnitude contrast ($\omega_{\text{m}}=20.6$~MHz, indicated by the gray dashed line) is shown in the lower panel, and the corresponding HN couplings are shown in \textbf{b}. Since the exceptional rings correspond to the zero crossings of one of the two couplings shown in \textbf{b}, the phase contrast inside the rings stays at $\pi$ as a result of the opposite signs of $\lambda_{12}$ and $\lambda_{21}$ (shown in \textbf{c}). On the other hand, the magnitude contrast is zero when $|\lambda_{12}|=|\lambda_{21}|$. 
 The only such points occur at $\phi = 0^\circ$ and $\pm180^\circ$, where there is no sign difference between the two couplings and thus no gyration.
 The situation with a stronger modulation strength ($\beta=26$~MHz here) differs qualitatively (\textbf{d}-\textbf{f}). In this case the two exceptional rings intersect, as shown in \textbf{d}. The intersection of the two rings combined with the continuity of the magnitude contrast force zeros in magnitude contrast at non-trivial $\phi$. A line-cut of magnitude contrast for $\omega_{\text{m}} = 20.6$~MHz (indicated by the gray dashed line in \textbf{d}) is shown in \textbf{d} bottom as an example. \protect\specialstar\ and \protect\specialtriangle\ highlight the two zero crossings. 
At these points $\lambda_{21}=-\lambda_{12}$ (\textbf{e}), resulting in pure gyration with both perfect $\pi$ phase contrast (\textbf{f}) and equal magnitudes.
}
	\label{Fig. S2}
\end{figure*}

\section{\label{sup:HN model}Equivalent Hatano-Nelson Model}
Although all the transmission properties can be now numerically solved with Eq. \ref{eq_S16}, we can rewrite Eq. \ref{eq_S7} in a more insightful way to show equivalence between the effective Hamiltonian of the zeroth sideband and a generalized Hatano-Nelson model \cite{Orsel2025}.

% \ak{inserting potential wording}
% The full theoretical description of how to obtain effctive HN coefficients from a single-sideband model were derived in Ref.~\cite{Orsel2025}, so we will not provide a full treatment here. We instead summarize the key aspects of the CPW resonator modes and coupling structure necessary to show that resonator lattices can give rise to two 

In the process of deriving Eq. \ref{eq_S16}, Eq. \ref{eq_S7} can be written as
\begin{equation}
\begin{aligned}
    \omega\ket{a(\omega)}&=[H_0+O_{\text{-P}}+\Tilde{O}_{\text{P}}]\ket{a(\omega)}-iK^T\ket{s_+(\omega)} \\
    &=H_\text{eff}\ket{a(\omega)}-iK^T\ket{s_+(\omega)},
\end{aligned}
\label{eq_S18}
\end{equation}
where all the contributions from the sidebands are mapped into the non-Hermitian effective Hamiltonian of the zeroth sideband. In this way, the non-reciprocity is imprinted in the off-diagonal asymmetric Hatano-Nelson couplings $\lambda_{12}$ and $\lambda_{21}$, as 
\begin{equation}
H_\text{eff} = \begin{pmatrix}
\omega_0+i\gamma_{\text{eff}} & \lambda_{12} \\
\lambda_{21} & \omega_0+i\gamma_{\text{eff}}
\end{pmatrix},
\label{eq_HN}
\end{equation}
where the total decay rate for each resonator changes from $\gamma$ to $\gamma_{\text{eff}}$ to compensate for the extra loss due to sideband conversion. By varying the three modulation parameters $\beta, \omega_\text{m}$, and $\phi$, it is possible to tune $\lambda_{12}$ and $\lambda_{21}$ continuously (see Fig. \ref{Fig. S2}b and e). If the diagonal entries of $H_\text{eff}$ are identical, the non-Hermitian system reaches an exceptional point (EP) when one of the couplings becomes zero. The corresponding transmission is then completely suppressed (see Fig. \ref{Fig. S2}a and d). After a zero crossing, the two couplings have opposite signs, giving rise to a $\pi$ phase contrast between forward and backward transmission (see Fig. \ref{Fig. S2}c and f).

\begin{figure*}[th!]
\centering
        \includegraphics{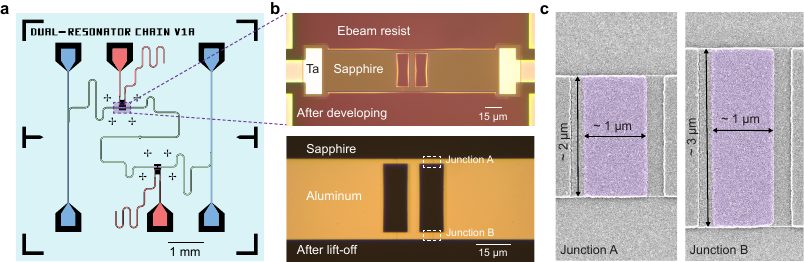}
	\caption{\textbf{CAD design, optical images, and SEM images of the device.} 
 \textbf{a}, Device layout in CAD. Identical transmission line resonators (green) are capacitively coupled. A modulator is introduced in each resonator using a dc-SQUID array (purple), controlled by on-chip flux bias lines (red). Two feedlines (blue) provide input and output ports. \textbf{b}, optical images of the SQUID area right after development (top), and after lift-off (bottom). \textbf{c}, False-color SEM images of the two asymmetric Josephson junctions (purple) in a dc-SQUID loop.
}
	\label{Fig. S3}
\end{figure*}

To prove the guaranteed intersections of the exceptional rings at higher modulation strength, we solve a simplified model analytically, considering only the first-order sidebands. %\gb{Previous sentence is a bit run on?} 
By setting the cut-off order $P=1$ in Eq. \ref{eq_S17} and deriving $H_{\text{eff}}$ from Eq. \ref{eq_S18}, the asymmetric couplings can be expressed as
\begin{equation}
\begin{aligned}
    \lambda_{12}&=\lambda(1-\alpha_{12}),\\
    \lambda_{21}&=\lambda(1-\alpha_{21}),
\end{aligned}
\label{eq_S19}
\end{equation}
where
\begin{equation}
\begin{aligned}
    \alpha_{12}&=\frac{\beta^2}{2}\frac{(\gamma_0^2+\lambda^2-\omega_{\text{m}}^2)\cos(\phi)+2\gamma_0\omega_{\text{m}}\sin(\phi)}{[\gamma_0^2+(\lambda-\omega_{\text{m}})^2][\gamma_0^2+(\lambda+\omega_{\text{m}})^2]},\\
    \alpha_{21}&=\frac{\beta^2}{2}\frac{(\gamma_0^2+\lambda^2-\omega_{\text{m}}^2)\cos(\phi)-2\gamma_0\omega_{\text{m}}\sin(\phi)}{[\gamma_0^2+(\lambda-\omega_{\text{m}})^2][\gamma_0^2+(\lambda+\omega_{\text{m}})^2]}.
\end{aligned}
\label{eq_S20}
\end{equation}
The coupling strength modification factors \(\alpha_{12}\) and \(\alpha_{21}\) contain a scaling coefficient, $\beta^2/2$; a symmetric term, proportional to \(\cos(\phi)\); and an asymmetric term, proportional to \(\sin(\phi)\). For any given $\beta$ and $\omega_\text{m}$, the difference between \(\lambda_{12}\) and \(\lambda_{21}\) is maximized when \(\phi=\pm90^\circ\) and vanishes at $\phi=0^\circ$ or $\pm 180^\circ$ (trivial $\phi$), where $\alpha_{12} = \alpha_{21}$.
The exceptional rings intersect whenever $\lambda_{21} = 0 = \lambda_{12}$, which requires $\sin(\phi) = 0$, and $\alpha_{21} = \alpha_{12} = 1$. Both of these conditions can be fulfilled at $\phi = 0$ for sufficiently large $\beta$ since $\alpha_{ij} \propto \beta^2$.

In reality, at large modulation strength, multiple orders of sidebands are populated, and the simplified analytical model discussed above is no longer accurate. Nevertheless, our numerical simulation results (Fig. \ref{Fig. S2}) show that the qualitative conclusions remain the same. The exact conditions required ($\beta$ and $\omega_\text{m}$) for the intersection will differ from what Eq. \ref{eq_S20} predicts, but the intersection itself must exist at $\phi=0$ when $\beta$ is sufficiently large.

% Numerical simulations of the non-reciprocal couplings and isolation and gyration behavior for realistic device parameters are shown in Fig.~\ref{fig:oldfig3}. For a drive strength of $\beta = XYZ$, the chain is still below the pure gyration threshold, and the EP surfaces do not intersect. However, by $\beta = XYZ$, in intersection emerges, which introduces a continuous set of pure gyration points.

% This suggests that for sufficiently large \(\beta\), \(\alpha_{21}=\alpha_{12}=1\) can always be achieved at $\phi=0$. Consequently, \(\lambda_{12}=\lambda_{21}=0\), which predicts the crossing of exceptional rings at $\phi=0$ as shown in Fig. 3d in the main manuscript.

% The coupling strength modification factors \(\alpha_{12}\) and \(\alpha_{21}\) contain a scaling coefficient, $\beta^2/2$, a symmetric term, proportional to \(\cos(\phi)\), and an asymmetric term, proportional to \(\sin(\phi)\). For any given $\beta$ and $\omega_\text{m}$, the difference between \(\lambda_{12}\) and \(\lambda_{21}\) is maximized when \(\phi=\pm\pi/2\) and vanishes at $\phi=0$ or $\pm \pi$. This suggests that for sufficiently large \(\beta\), \(\alpha_{21}=\alpha_{12}=1\) can always be achieved at $\phi=0$. Consequently, \(\lambda_{12}=\lambda_{21}=0\), which predicts the crossing of exceptional rings at $\phi=0$ as shown in Fig. 3d in the main manuscript.

\begin{figure}[!htbp]
\centering
        \includegraphics{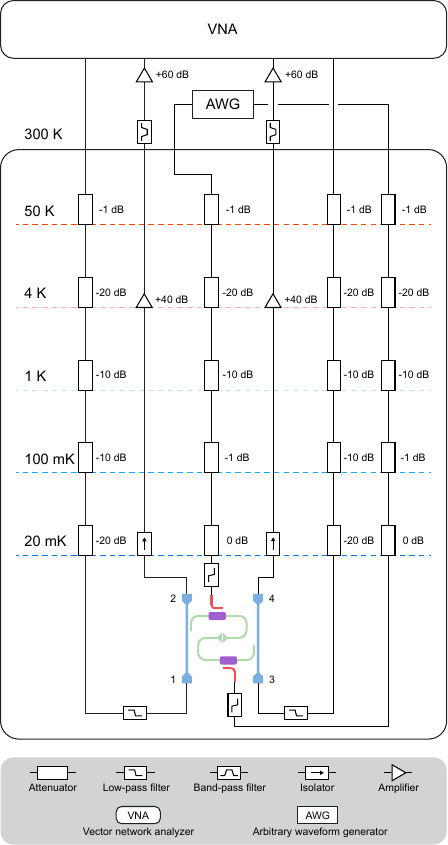}
	\caption{\textbf{Experimental setup.} Schematic of the wiring within the dilution refrigerator and the equipment setup outside. Two input lines (to ports 1 and 3), two output lines (from ports 2 and 4), and two flux bias lines (red) are shown in the diagram.
}
	\label{Fig. S4}
\end{figure}

\begin{figure*}[!htbp]
\centering
        \includegraphics{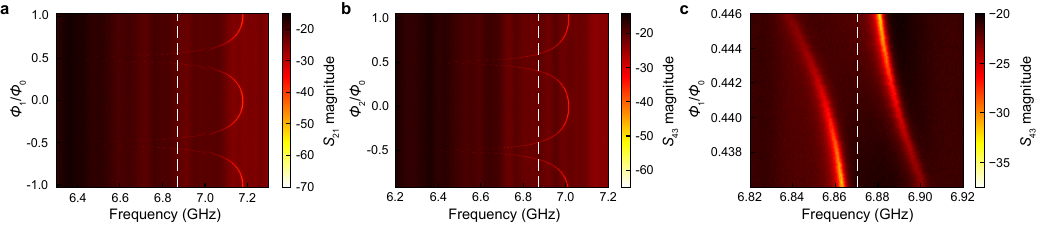}
	\caption{\textbf{Flux dependence of resonance frequencies and the extraction of \boldmath$\lambda$.} 
 \textbf{a,b}, Resonance frequencies (indicated by the low transmission magnitudes in $S_{21}$ and $S_{43}$ measurements) as a function of the applied magnetic flux for resonator 1 and 2, respectively. $\Phi_0=h/2e$ is the superconducting magnetic flux quantum. The white dashed lines indicate the center frequency $\omega_0=6.8705$~GHz chosen for the remainder of the results, similarly for \textbf{c}. The avoided crossings observed in \textbf{a} around $7$~GHz are due to the coupling to resonator $2$, whose resonance frequency is $\sim7$~GHz with zero applied magnetic flux. \textbf{c}, Avoided crossing between the two resonators near $\omega_0$. The data is acquired by fixing resonator 2 at $\omega_0$ while varying the applied magnetic flux on resonator 1 to drive it across resonator 2. The inter-resonator coupling strength $\lambda$ is determined to be $16.4$~MHz based on the minimum spacing of $2\lambda$.
}
	\label{Fig. S5}
\end{figure*}

\begin{table*}[!htbp]
\caption{Resonator Characterization} 
\centering % used for centering table
\begin{tabular}{c c c c c c c c} % centered columns 
\hline \\[-2.5ex]
Resonator index & $Q_\text{int}$ & $Q_\text{ext}$ & Linewidth & Port couplings & $\kappa$ & $\gamma$\\  % inserts table
%heading
\hline \\[-2ex]
1 & 17484 & 923 & 7.8~MHz & $k_1^2,k_2^2\approx3.7$~MHz & 0.4~MHz & 3.9~MHz\\ % inserting body of the table
2 & 17271 & 1000 & 7.3~MHz & $k_3^2,k_4^2\approx3.4$~MHz & 0.4~MHz & 3.6~MHz\\[0.4ex]
\hline 
\end{tabular}
\label{table_1}
\end{table*}

\begin{figure*}[!htbp]
\centering
        \includegraphics{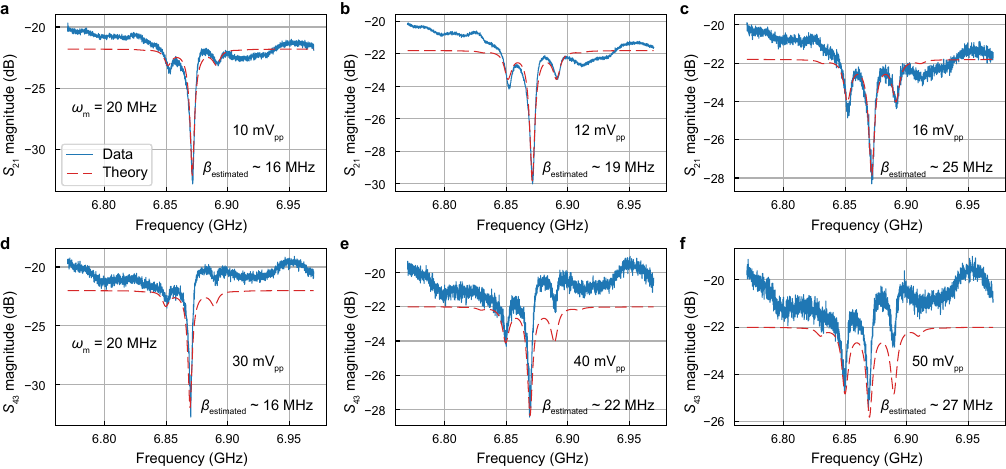}
	\caption{\textbf{Calibration of \boldmath$\beta$.} 
 \textbf{a-c}, Comparison between measured $S_{21}$ magnitude and numerical simulations under different modulation strengths for resonator 1. The extracted ratio of $\beta$ (MHz)/applied ac voltage (m$\text{V}_{\text{pp}}$) is $\sim1.58$. The slight mismatch between the theory and data mainly comes from the non-flat baseline due to line imperfections, and the non-linearity in the flux dependence of the resonance frequency as shown in Fig. \ref{Fig. S5} (similarly for \textbf{d-f}). \textbf{d-f}, Comparison between measured $S_{43}$ magnitude and numerical simulations under different modulation strengths for resonator 2. The extracted ratio of $\beta$ (MHz)/applied ac voltage (m$\text{V}_{\text{pp}}$) is $\sim0.54$.
}
	\label{Fig. S6}
\end{figure*}

\section{\label{sup:fab}Device Fabrication and SQUID Design}

The device is fabricated on a $7\times7$ $\text{mm}^2$ sapphire substrate with a tantalum (Ta) superconducting top layer. Coplanar waveguide (CPW) transmission line resonators, feedlines, and the flux bias lines are patterned using standard photolithography followed by wet etching (CAD layout shown in Fig. \ref{Fig. S3}a), while the SQUID arrays are fabricated by electron-beam lithography and shadow evaporation of aluminum (optical and SEM images shown in Fig. \ref{Fig. S3}b,c). Each array consists of two $11\times30$ $\mu\text{m}^2$ dc-SQUID loops, within which the two Dolan-style Josephson junctions are designed to have areas of $2 \, \mu\text{m}^2$ and $3 \, \mu\text{m}^2$, respectively. The Josephson energies are estimated to be $E_\text{J1}/h=2$ THz and $E_\text{J1}/h=3$ THz, corresponding to critical currents of $I_\text{c1}=4 \, \mu$A and $I_\text{c2}=6 \, \mu$A, respectively (critical current density: $\sim2 \, \mu A/\mu\text{m}^2$). Based on Eq. 1 in the main manuscript, the tuning range of the total effective inductance from one SQUID array is then estimated to be $0.066$\textendash$0.329$ nH.

\section{\label{sup:abs}Measurement Setup and Calibrations}

The experimental setup used throughout the measurements in the main manuscript is depicted in Fig. \ref{Fig. S4}. The device is mounted on the base plate of the dilution refrigerator and cooled down to $20$ mK. Ports 1 and 3 of the device are connected to two input lines (with attenuators for isolation from thermal noise at higher temperature stages), and ports 2 and 4 are connected to two output lines (with isolators and amplifiers). We perform all the transmission measurements using a Rohde \& Schwarz ZNB20 vector network analyzer (VNA). We apply a small sinusoidal ac bias on top of a dc offset through the flux bias line 
(similar to input lines but with low-pass filters that have a much lower cut-off frequency, and less attenuation) to modulate each SQUID array, using a dual-channel Keysight 33622A arbitrary waveform generator (AWG). The differential phase $\phi$ between the two modulation tones is controlled by setting the phase offset between the two channels within the AWG. 

\subsection{Individual resonator characterization}
The flux dependence of the resonance frequency of each resonator is characterized by single-sided transmission measurements ($S_{21}$ and $S_{43}$) at different dc bias voltages, as shown in Fig. \ref{Fig. S5}a and b. Both resonators show periodic behavior of their resonance frequencies as expected from the dependence of the dc-SQUID inductance on external magnetic flux (described by Eq. 1 in the main manuscript). The nearly perfect periodicity also suggests a good homogeneity of the magnetic field threading the different SQUID loops. As a result, resonator 1 has a tuning range from $6.6$ to $7.2$ GHz and resonator 2 tunes from $6.4$ to $7$ GHz. 

We choose $\omega_0=6.8705$~GHz as the center frequency for our further experiments. The $Q$ factors, loss rates, and the extracted couplings between ports and resonators are listed in Table \ref{table_1}. The inter-resonator coupling strength $\lambda$ between two resonators is extracted from the minimum spacing ($2\lambda$) of the two hybridized modes when the resonance frequencies of the two resonators cross each other, as shown in Fig. \ref{Fig. S5}c. The coupling strength $\lambda$ extracted from this avoided crossing at $\omega_0$ is $16.4$~MHz.

\begin{figure}[!htbp]
\centering
        \includegraphics{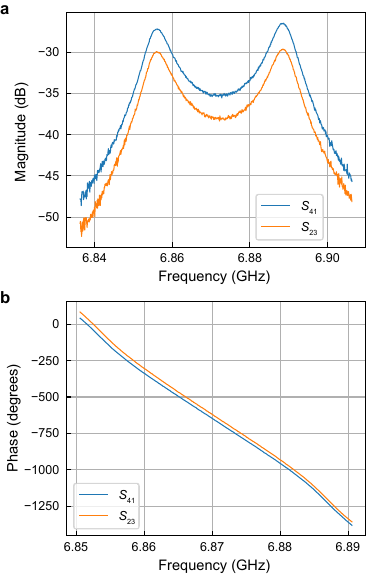}
	\caption{\textbf{Differential line loss and phase delay calibrations.} All the data shown here are taken under zero modulation, which means the chain is fully reciprocal and any offset originates outside the chain. \textbf{a}, Raw traces of $S_{41}$ and $S_{23}$ magnitudes. The line loss offset ($\sim3$~dB) between the two transmission curves is independent of the presence of modulation, and is compensated for in all magnitude contrast measurements. The same method applies to phase contrast measurements, except that the line delay offset exhibits frequency dependence (shown in \textbf{b}). As a result, a frequency-dependent compensation is adopted. \textbf{b}, Raw traces of $S_{41}$ and $S_{23}$ unwrapped phases.}
	\label{Fig. S7}
\end{figure}

\subsection{\texorpdfstring{Calibration of \boldmath$\beta$}{beta}}
To determine the modulation strength applied to each resonator, it is essential to extract the conversion ratio between $\beta$ and the applied ac bias voltage. Note that it is unreliable to extract this ratio from the flux dependence of the resonance frequency shown in Fig. \ref{Fig. S5}, since the attenuations on the flux bias lines are different for dc and ac currents. The ability to probe each resonator through $S_{21}$ and $S_{43}$ even when both resonators are far detuned and direct transmission through the chain ($S_{41}$ and $S_{23}$) is zero allows the modulation parameters for each resonator to be calibrated fully independently. To extract $\beta$, we compare the measured and calculated transmission results (using the method described in {\S}\ref{sup:input-output}) under multiple ac bias voltages. As shown in Fig. \ref{Fig. S6}a-c, a conversion ratio of $\sim1.58$ MHz/m$\text{V}_{\text{pp}}$ aligns the simulations with experiments well for resonator 1. Similarly, a ratio of $\sim0.54$ MHz/m$\text{V}_{\text{pp}}$ is extracted for resonator 2 in Fig. \ref{Fig. S6}d-f. Note that, the finite discrepancy between the simulations and experiments mainly comes from the non-flat baseline due to line imperfections, and the non-linearity in the flux dependence of the resonance frequency as shown in Fig. \ref{Fig. S5}. Ultimately, this calibration method serves as an estimation of $\beta$, and as we have shown in the main manuscript, observing pure gyration does not require fine tuning of $\beta$.

\subsection{Line loss and phase delay calibrations}

In order to extract the true magnitude contrast and phase contrast generated by the modulated resonator chain, a fair comparison between $S_{41}$ and $S_{23}$ is needed, which requires being able to remove the offsets within the transmission lines and cables outside the chain. Thanks to our unique 4-port design, this can be done in situ by using reference traces of $S_{41}$ and $S_{23}$ when no modulation is applied (see Fig. \ref{Fig. S7}). The resonator chain itself is fully reciprocal without the spatio-temporal modulation, hence any offset detected during this measurement stems from the differential attenuation and phase delay outside the chain. The line loss offset ($\sim3$~dB) between the two transmission curves (shown in Fig. \ref{Fig. S7}a), which is independent of the presence of modulation, is compensated for in all magnitude contrast measurements. The same method applies to phase contrast measurements, except that the line delay offset exhibits frequency dependence (shown in Fig. \ref{Fig. S7}b). As a result, a frequency-dependent compensation is adopted. On top of that, the insertion loss of the device (magnitude data in Fig. 2b in the main manuscript) is calibrated by comparing the calculated peak magnitudes of the two hybridized modes (using method in {\S}\ref{sup:input-output}) with data measured under zero modulation (Fig. \ref{Fig. S7}a). A nearly $-20$~dB round-trip loss outside the device was found and was crosschecked by another cryogenic measurement where the device was replaced by a short SMA cable.

Lastly, we calibrate the differential phase $\phi$ between the two modulation tones. Given the slight difference in cable lengths/filter responses on the two flux bias lines, the two ac modulation signals will gain an extra unknown phase offset by the time they reach the device at the bottom of the dilution refrigerator. However, this offset can be detected through a $\phi$ sweep (see Fig. 2c in the main manuscript) since the non-reciprocal behavior of the system is anti-symmetrical around $\phi=0$. In this way, a $\sim5.4^\circ$ extra phase offset (for $\omega_\text{m}\sim20$~MHz) between the two flux bias lines is extracted and used for $\phi$ calibration in Fig. 2 and Fig. 3 in the main manuscript. As a reference, the wavelength of a $20$~MHz sinusoidal signal in a coax cable is $\sim10$ m, hence a $5.4^\circ$ phase delay corresponds to a $\sim0.15$-m difference in cable lengths.

%%%%%%%%%%%%%%%%%%%%%%%%%%%%%%%%%%%
\bibliographystyle{apsrev4-2}
\bibliography{sup_refs}